\documentclass[preprint, aps,showpacs]{revtex4}
\usepackage{graphicx}
\usepackage{amsmath}
\usepackage{amssymb}
\usepackage{hyperref}
\usepackage{latexsym}
\usepackage{epsfig}
\usepackage{array}


\begin{document}

\title{Enhancement of mechanical Q-factors  by optical trapping}
\author{K.-K. Ni$^{1}$}
\author{R. Norte$^{2}$, D. J. Wilson$^{1}$, J. D. Hood$^{1}$}
\author{ D. E. Chang$^{1, 3}$}
\author{ O. Painter$^{2}$}
\author{ H. J. Kimble$^{1}$}
\affiliation{$^{1}$Norman Bridge Laboratory of Physics 12-33, California Institute of Technology,
Pasadena, CA 91125}
\affiliation{$^{2}$Thomas J. Watson, Sr., Laboratory of Applied Physics, California Institute of Technology, Pasadena, CA 91125}
\affiliation{$^{3}$ICFO - Institut de Ciences Fotoniques, Mediterranean Technology Park, 08860 Castelldefels, Barcelona, Spain}



\date{\today}

\begin{abstract}

The quality factor of a mechanical resonator is an important
figure of merit for various sensing applications and for
observing quantum behavior. 
Here, we demonstrate a technique to push the quality factor
of a micro-mechanical resonator beyond conventional material and fabrication limits by
using an optical field to stiffen or ``trap'' a particular
motional mode. Optical forces increase the
oscillation frequency by storing most of the mechanical energy in a
lossless optical potential, thereby strongly diluting the effect of
material dissipation. By using a 130 nm
thick SiO$_2$ disk as a suspended pendulum, we 
achieve an increase in the pendulum center-of-mass
frequency from 6.2 kHz to 145 kHz. The corresponding quality
factor increases 50-fold from its intrinsic value to a final value of
$Q=5.8(1.1)\times 10^5$, representing more than an order of magnitude 
improvement over the conventional  limits of SiO$_2$ for this geometry. 
Our technique may enable new opportunities for
mechanical sensing and facilitate observations of quantum behavior
in this class of mechanical systems.


\end{abstract}


\maketitle

Mechanical resonators are widely used as exquisite sensors
of weak perturbations such as small forces~\cite{hoyle01,mamin01},
displacements~\cite{abramovici92,teufel09}, and changes in
mass~\cite{lavrik03,ono03}. In fact, a number of systems have
advanced to the point that it is possible to detect quantum
effects in their
motion~\cite{teufel09,roucheleau10,oconnell10,chan11}, raising the
exciting possibility that such systems might eventually lead to
applications in quantum information processing~\cite{Stannigel2010, Chang2011b, Regal2011}
and the observation of quantum effects at macroscopic
scales~\cite{genes08b,muller-eberhardt08}.

The performance of a mechanical resonator depends critically on its
quality factor, which characterizes both the maximum response of
an oscillator to a disturbance at its resonance frequency and the
coupling rate to its surrounding dissipative environment.
Improving quality factors beyond state-of-the-art parameters is a
challenging task since  a number of systems are now limited
by fundamental dissipation mechanisms, e.g. thermoelastic 
damping~\cite{lifshitz00} and internal friction~\cite{pohl02}.

In this Letter, we experimentally demonstrate a technique
that enables the quality factor of a mechanical system to be
enhanced beyond conventional material limits. Our technique
involves optically trapping a thin, dielectric membrane whose
geometry is designed so that the natural material forces are
extremely weak~\cite{chang11}. In this limit, almost all
mechanical energy is stored in an ultralow loss potential provided by
strong optical restoring forces, which dilute the effects of
internal material dissipation~\cite{chang11, Cagnoli2000}. 
Our general scheme is implemented for a particular example of an
SiO$_2$ dielectric disk supported by a single thin tether,
trapped in an optical standing wave. We observe an
increase in the ``pendulum'' mode frequency from $6.2$~kHz to
$145$~kHz as the optical power is increased, leading to a final
quality factor of $Q_f=5.8(1.1)\times 10^5$. $Q_f$ represents
greater than fifty-fold increase over the intrinsic $Q_i$ of our device in the absence
of optical trapping forces, and significantly, more than an order of magnitude improvement
over estimates of the conventional dissipative rate of our
SiO$_2$ disk~\cite{Yasumura2000, Penn2006}. These results substantiate
the potential of our technique to facilitate mechanical sensors with enhanced 
sensitivity and quantum devices based upon mechanical systems.

\begin{figure}[!t]
    \includegraphics[width=\columnwidth]{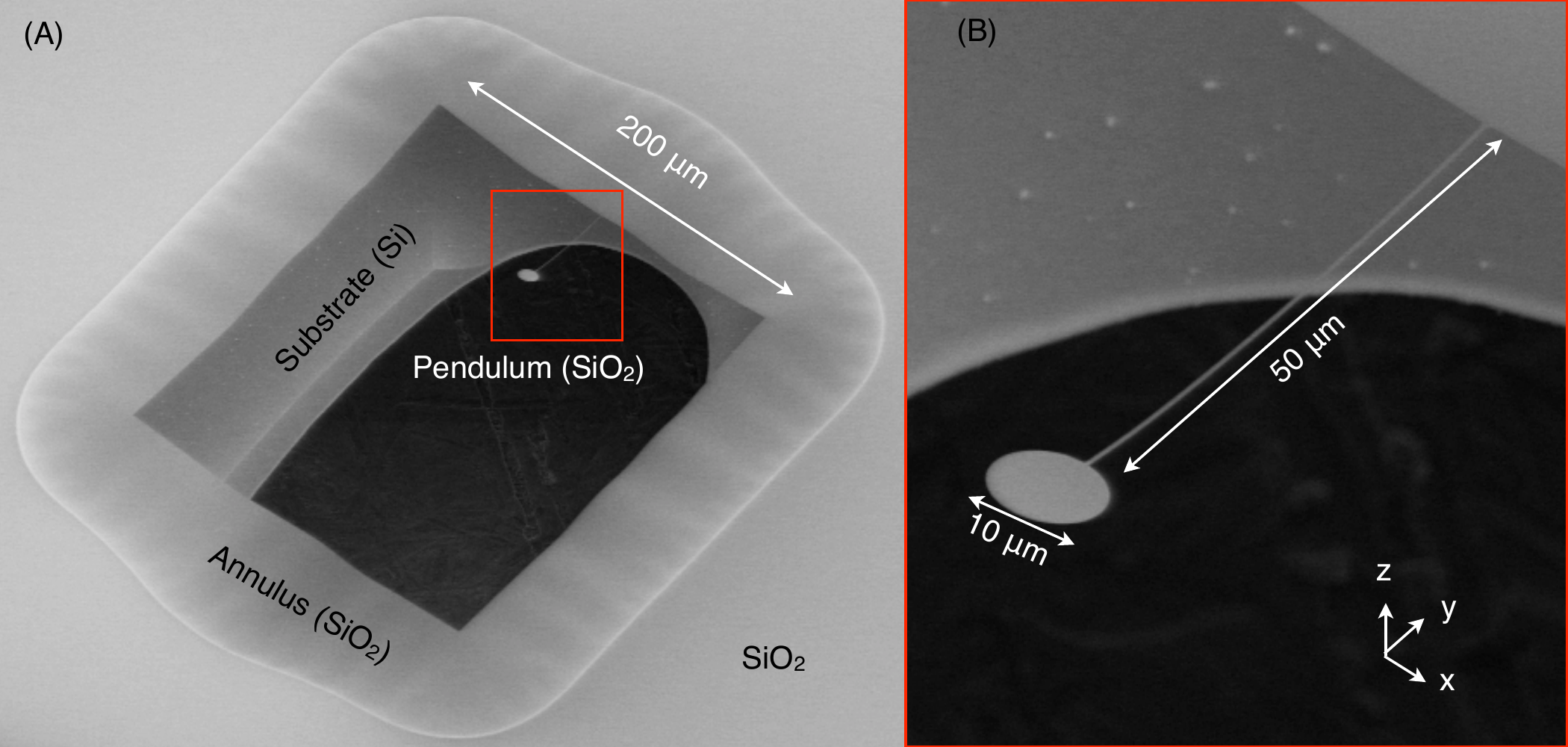}
  \caption{Scanning electron micrographs of the device:  a 130 nm thick SiO$_2$ membrane forms the pendulum, which consists of a 10 $\mu$m diameter disk and a 50 $\mu$m $\times$ 0.43 $\mu$m tether. (A) Overview: The pendulum is suspended from a 60 $\mu$m wide SiO$_2$ annulus.  The annulus (wrinkled area) and the pendulum are etched into a SiO$_2$ film for which the Si substrate directly underneath has been removed. The dark background in the center is a clear opening of the substrate.  (B) Close-up view of the pendulum, which is deflected 10-15 degrees out of the plane of the substrate due to residual stress of the film.}
\label{device}
\end{figure}

Optical forces are generally feeble as compared to mechanical forces, which makes optical manipulation of mechanical oscillators challenging.
To implement optical trapping of membranes, we begin by fabricating a nearly free-standing dielectric film in a pendulum geometry. We chose SiO$_2$ as the membrane material mainly for its low optical absorption~\cite{Armani03}. The pendulum (Fig. \ref{device}) consists of a 10 $\mu$m diameter disk held by a 50 $\mu$m $\times$ 0.43 $\mu$m tether, which is attached to a large, square annulus of SiO$_2$ that has a width of $\sim$ 60 $\mu$m. The thickness of all the suspended parts (i.e. the disk, the tether, and the annulus) is 130 nm. 

The pendulum mechanical resonators are fabricated from a 200 $\mu$m thick Si wafer in which a 130 nm surface SiO$_2$ layer has been formed using dry oxidation.
Electron beam lithography, followed by a C$_4$F$_8$:SF$_6$ plasma etch, are used to transfer the disk and tether pattern into the surface SiO$_2$ layer of 1$\times$1 cm chips diced from the wafer.  A XeF$_2$ plasma etch is used to selectively remove the underlying Si, releasing the SiO$_2$ pendulum and opening a through-hole in the Si substrate.  Due to the isotropic nature of the XeF$_2$ etch, a narrow annulus of undercut SiO$_2$ is formed at the periphery of the transferred pattern (see Fig. \ref{device}).  To reduce the width of the SiO$_2$ annulus, the Si substrate below the pendulum pattern is pre-thinned from the backside.

\begin{figure}[!t]
    \includegraphics[width=\columnwidth]{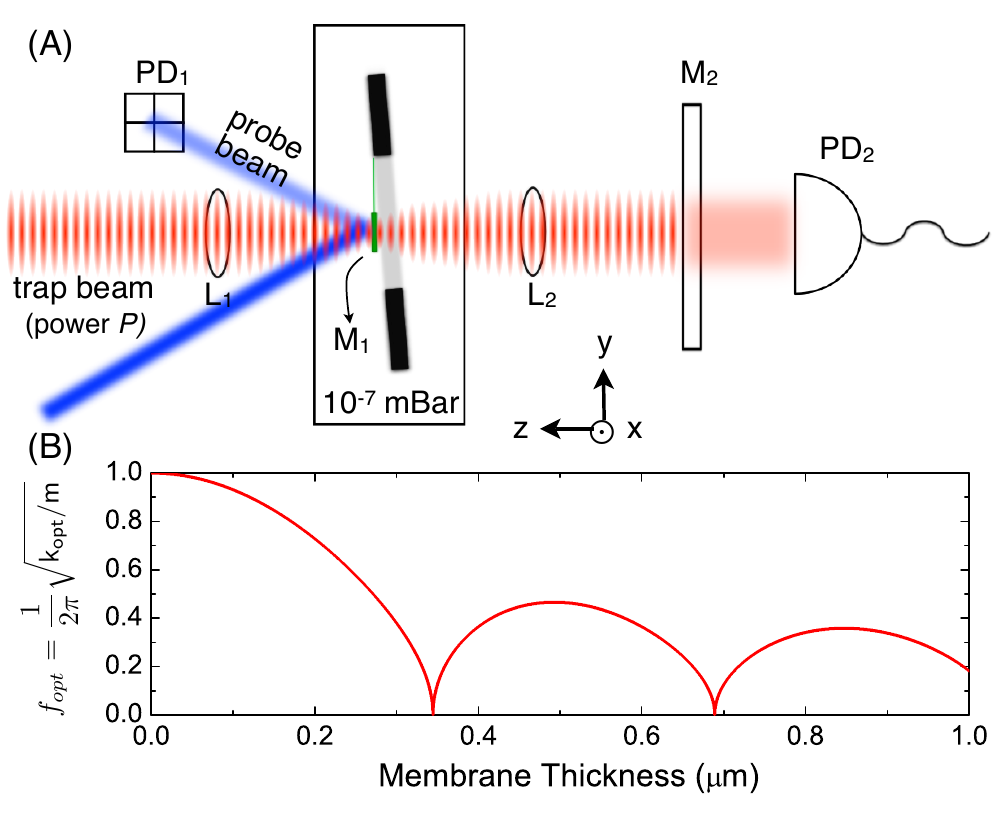}
  \caption{Optical trapping of a membrane disk. (A) Schematic of the experimental setup.  The membrane chip is enclosed inside a vacuum chamber. We trap the disk in an optical standing wave formed by a single laser beam at 1.064 $\mu$m and its reflection from the disk (M$_1$) and a mirror (M$_2$, reflectivity = 0.98).  We monitor the thermal motion of the pendulum by the deflection of an off-axis probe beam (blue) reflected from the disk onto a quadrant photodiode (PD$_1$) and transmitted intensity onto PD$_2$.
(B) Calculated trap frequency $ f_{opt}$ vs membrane thickness $d_m$ normalized to $f_{opt}$ at $d_m/\lambda \rightarrow 0$ for a fixed optical power.  
For $d_m \sim \lambda$, we solve for the steady electric-field amplitude on the left and right of the membrane. Variation of the standing-wave amplitude within the membrane leads to variation of $f_{opt}$ at its equilibrium position. }
\label{setup}
\end{figure}

To investigate properties related to a trapped pendulum, we load the 1$\times$1 cm pendulum chip, which typically contains a dozen devices, into a vacuum chamber traversed by an optical standing wave. A schematic of the experimental setup is shown in Fig. \ref{setup}(A). The pendulums hang vertically inside the chamber, which is evacuated to a pressure below $10^{-7}$ mbar to make gas damping negligible.
The optical standing wave is formed by a retro-reflected Gaussian beam, which has been focused to a $1/e^2$ waist $\omega_0\simeq 17$ $\mu$m at the position of the disk. The trap beam is derived from a high power Nd:YAG laser operating at a wavelength of $\lambda=$1.064 $\mu$m. We vary the incident laser power between 3 mW to 17 W using a waveplate and a polarizing cube. The centering of the trap beam on the pendulum disk is critical for achieving large trapping potentials without mixing ``center-of-mass'' (CM) motion of the pendulum with vibrational modes of the tether.  This degree of freedom is carefully aligned by monitoring transmission of the forward and retro-reflected beams through the disk. In addition, to ensure that the disk is perpendicular to the optical standing wave, we implement a pair of actuators to tip and tilt the chip to prevent the pendulum from settling into a configuration in which the disk extends over multiple periods of the standing wave. 

For the optical standing wave configuration shown schematically in Fig. \ref{setup}(A), we estimate the trapping potential by balancing expressions for the radiation pressure force on either side of the membrane. To simplify the problem, we assume that the disk is infinitely stiff, that it is free to move along the axis of the trap, and that the incident optical beam is smaller than the disk so that diffraction from the edges may be ignored. We solve for the electric-field of a single beam in the presence of two reflectors, M$_1$ and M$_2$ (where M$_1$ represents the disk), as a function of the membrane thickness, $d_m$. Stable equilibria occur at the positions where
the force between the incoming beam and the finite electric field built up between M$_1$ and M$_2$ are balanced, which are neither at nodes nor anti-nodes of the standing wave.
At each equilibrium position, the optical spring constant of the trap is $k_{opt} = \frac{16\pi}{\lambda}\frac{\left|r_m\right|}{\left|t_m\right|}\frac{P}{c}$, where $P$ is the incoming power, $r_m$ and $t_m$ are the reflectance and the transmittance of the membrane determined by $d_m$ and the index of refraction of the film~\cite{Hecht}, and $c$ is the speed of light. In Fig. \ref{setup}(B), we plot the calculated optical trapping frequency, $f_{opt}=\frac{\sqrt{k_{opt}/m}}{2\pi}$, normalized to the prediction for a membrane with $d_m/\lambda \rightarrow 0$. For our membrane with thickness $d_m=0.13 \,\mu$m, the trapping frequency is 88\% of the value predicted for a membrane with $d_m/\lambda \rightarrow 0$.
It is interesting to note that the optical spring constant for a membrane in the middle of a high finesse Fabry-Perot cavity is also  $k_{opt} = \frac{16\pi}{\lambda}\frac{\left|r_m\right|}{\left|t_m\right|}\frac{P}{c}$, where $P$ is now the circulating power. In this case, however, the membrane center would be trapped specifically at either an anti-node or a node of the standing wave depending on the membrane thickness. 



\begin{figure}[!t]
    \includegraphics[width=5.5in]{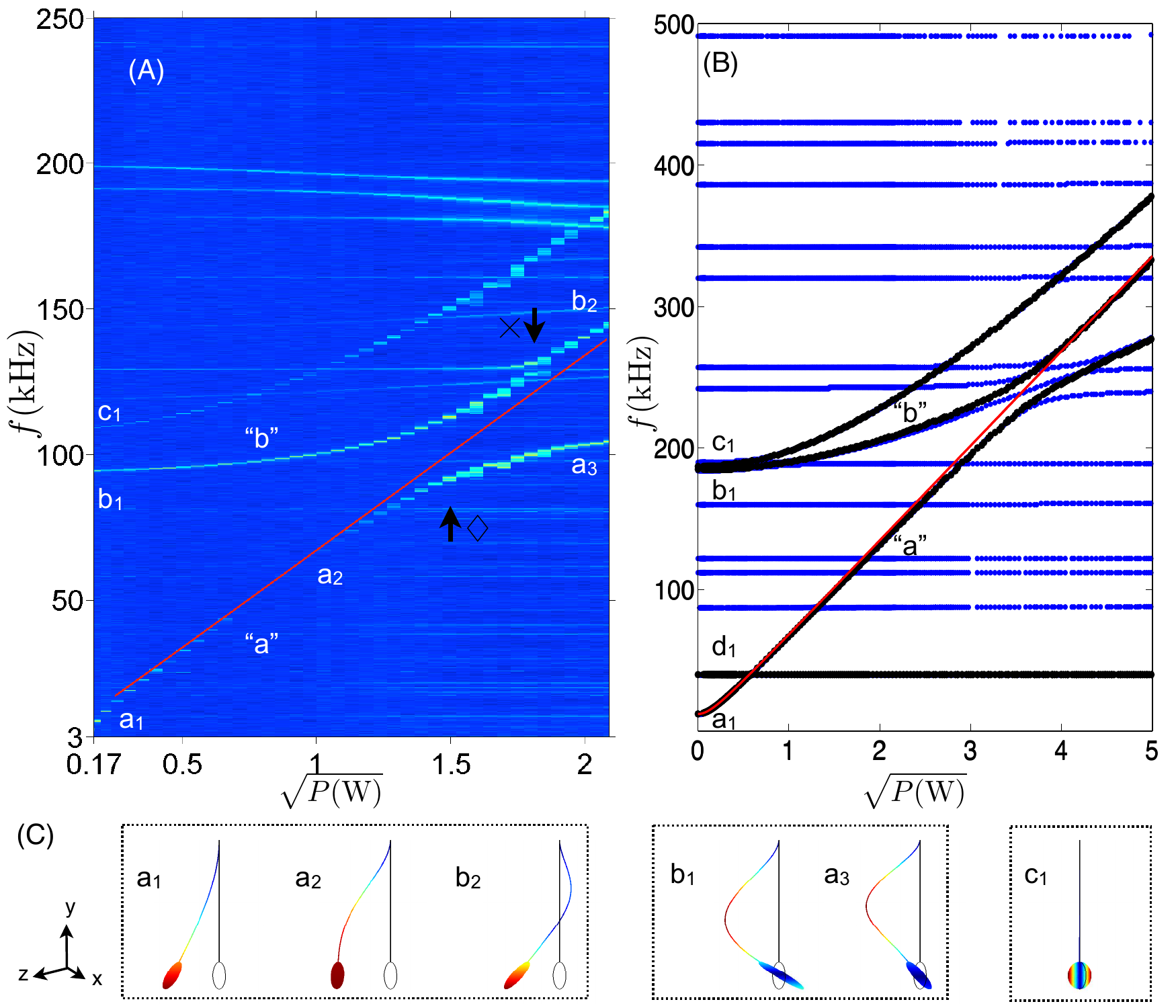}
  \caption{Displacement power spectrum of vibrational modes. (A) Spectrogram of vertical angular displacement (inferred from deflection measurement, Fig. 2(A)) versus trapping power, $P$.  Several vibrational mode branches are evident, e.g. ``a" and ``b".  
  As the optical trap power is increased (in discrete steps), the frequencies of the three lowest modes (a$_1$, b$_1$, c$_1$) increase as $\sqrt{f_{0i}^2+\alpha_{i} \cdot P}$ (see text). Two avoided crossings are visible here: ($\diamondsuit$) formed by the CM mode and the violin mode; ($\times$) formed by the CM mode and an annulus mode. At higher powers, the CM mode is a hybrid of pendulum and violin motion (b$_2$). Other visible modes in the spectrum are associated with vibrations of the annulus.  Overall, the CM frequency shifts from 6.2 kHz to 145 kHz when 4.3 W of optical power is applied.
(B) Finite-element simulated spectrum  of an optically trapped pendulum (black) suspended from an annulus (blue)  that is in turn anchored to a substrate (as in Fig. \ref{device}(a)) for qualitative comparison to (A). The red lines in both (A) and (B) are drawn for a trapping slope of  $\alpha = 4500\, \frac{\text{kHz}^{2}}{\text{W}}$.
(C)  Simulated mode shapes for different optical trapping forces in (A). (a$_1$) the ``pendulum" mode, also called the ``CM" mode, (b$_1$) the ``violin" mode, and (c$_1$) the ``torsional" mode. }
\label{spring}
\end{figure}

Intrinsic vibrational modes and frequencies of the pendulum structure are probed by reflecting an independent ``probe'' beam ($\lambda = 0.852$ $\mu$m) from the disk at an oblique angle (50$^{\circ}$) with respect to the trap.  The reflected beam is directed to a quadrant photodiode (PD$_1$)~\cite{Usami11} which is calibrated using the membrane tip/tilt actuators.
The quadrant photodiode can be split into two horizontal pairs of sensors (``top'' and ``bottom'') and two vertical pairs of sensors (``left'' and ``right'').  Analysis is performed on the difference between the ``top'' and ``bottom'' combined photosignals (expressed as a transimpedance-amplified photocurrent, $V_{TB}(t)$) and the difference between the ``left'' and ``right'' combined signals ($V_{LR}(t)$).  Fourier transforms of $V_{TB}(t)$ and $V_{LR}(t)$ reveal the frequencies and tip/tilt orientation of the vibrational modes (Fig. \ref{spring}(A)).  We identify low-order vibrational modes by comparison of the observed frequency spectrum and their characteristic mode shapes to  a finite-element simulation (Fig. \ref{spring}(B), COMSOL 3.5a) based on the membrane material properties and its geometry measured using a scanning electron microscope. 
We see approximately a factor of two discrepancy in the simulated frequencies versus measurements in the absence of optical forces. This is most likely due to an overestimate of the Young's modulus for the very thin SiO$_2$ layer where surface effects can be important.
The mode shapes in the absence of optical forces calculated from simulation are illustrated in Fig. \ref{spring}(C) as (a$_1$) the ``pendulum'' mode, also called the ``CM'' mode, where the pendulum disk  swings along the axis of the trapping beam ($f_{0a}=6.2$ kHz), (b$_1$) the ``violin'' mode ($f_{0b}=93$ kHz), and (c$_1$) the ``torsional'' mode ($f_{0c}=109$ kHz).  One additional mode that is not shown in the figure is (d$_1$) the transverse pendulum mode swinging orthogonal to the axis of the trapping beam (25 kHz).
The mode that we are most interested is the CM mode because it 
exhibits the least mechanical deformation which leads to energy dissipation~\cite{chang11}.

In the presence of an optical trap, the frequency of the membrane is determined by the sum of the optical restoring force and the intrinsic mechanical restoring force. The contribution from gravity is small ($<$ 100 Hz) and is generally neglected. The trap is first aligned at a low trapping power (3 mW).  
To ensure that the tether does not provide a significant initial restoring force, we also fine tune the trapping laser wavelength so that the equilibrium position of the trap coincides with the natural axial position of the membrane. We diagnose their coincidence by minimizing the vertical deflection of the probe beam (evident in the mean value of $V_{TB}$) when the trap is turned on.
Figure \ref{spring}(A) shows the power spectrum of $V_{TB}$, which reflects the vertical angular displacement of the pendulum, as a function of trap power, $P$. 
For comparison, we show a spectrum generated by a finite element model (Fig. \ref{spring}(B)) that treats the optical trap as a restoring force with a Gaussian transverse profile.  Close comparison of the observed and model spectra adds to our understanding of several distinct features.
As the optical trap power is increased, the frequencies of the three lowest visible modes increase as $\sqrt{f_{0i}^2+\alpha_{i} \cdot P}$, where $f_{0i}$ represents their natural frequency and $\alpha_{i}$ is the trapping slope coefficient for each mode. As the CM mode frequency increases, the mode shape also changes via the bending of the tether. This change is first evident in the reduction of the CM mode signal near 50 - 75 kHz. The reduction occurs as the mode shape changes from Fig. \ref{spring}(C, a$_1$) to (C, a$_2$), which to first order does not deflect vertically. In addition, as expected, we do not see such a reduction for corresponding trapping beam transmission on PD$_2$ (not displayed) that probes pure axial disk displacement. The mode shape again changes as the CM mode and the violin mode (Fig. \ref{spring}(C, b$_1$)) form an avoided crossing near 93 kHz. At higher power, the CM mode is a hybrid of pendulum and violin modes (Fig. \ref{spring}(C, b$_2$)).  As the frequency continues to increase, the CM mode crosses an annulus mode. Overall, the CM mode frequency shifts from 6.2 kHz to 145 kHz when 4.3 W of optical power is applied, corresponding to a trapping slope coefficient of $\alpha_{CM} = 4880\, \frac{\text{kHz}^{2}}{\text{W}}$, in good agreement with the finite-element simulated value of \ $\alpha_{CM}^{FEM} = 4500\, \frac{\text{kHz}^{2}}{\text{W}}$ (the inferred value from measurement being overestimated due to the mode anti-crossing). 
At trapping powers greater than 4.3 W, we find that the CM frequency ($f=$145 kHz) is near yet another vibrational mode of the annulus. With further increases in power, the large thermal displacements of the annulus mode greatly dominates the motion of the CM pendulum mode, making it difficult to identify. 

\begin{figure}[!t]
    \includegraphics[width=\columnwidth]{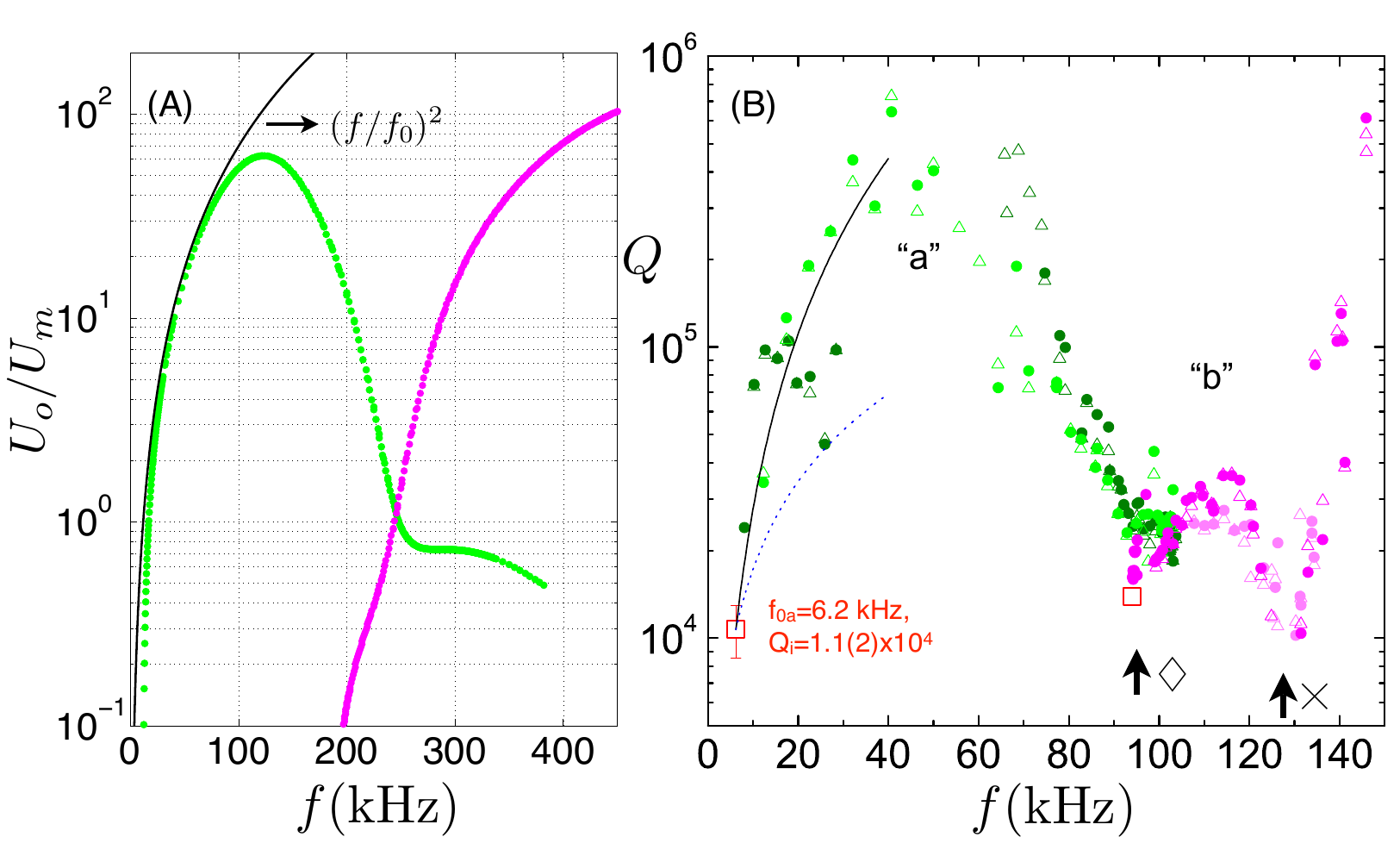}
  \caption{Mechanical $Q$-factor of the trapped pendulum. (A) Finite-element calculation of the ratio between the energy stored in the optical potential ($U_{o}$) and the mechanical potential ($U_{m}$) vs trap frequency. The two branches corresponding to ``a'' (green) and ``b'' (magenta) vibrational modes respectively in Fig. \ref{spring}.  (B) $Q$ vs trapping frequency for two vibrational modes in the trap, where $Q$ is inferred from the autocorrelation of the displacement energy, independently recorded on PD$_1$ (circles) and PD$_2$ (triangles). In addition, two dataset with slightly different optical trapping alignments are shown for ``a" (light and dark green) and ``b'' (light and dark magenta).    
  The $Q$ of the CM mode increases 60-fold from its natural value, $Q_i$ ($P=0$, data in red square) with a slope that is consistent with $(f/f_0)^2$ scaling (solid line) before turning over near the avoided crossing ($\diamondsuit$) with a violin mode. Beyond this avoided crossing, $Q$ increases again followed by a minimum near the annulus avoided-crossing ($\times$) and then a steep increase. We also show the expectation of a $Q$-increase that scales as $f/f_0$ (dashed line) for comparison.
 We find the measured $Q$ increase to be in qualitative agreement with the calculated $1+U_o/U_m$ in (A).
  }
\label{Qdata}
\end{figure}

One of the most important consequences of optical trapping is an increase of the mechanical $Q$~\cite{chang11}.  The large increase in frequency of the CM mode through optical trapping implies that the amount of mechanical energy stored in the optical fields $U_{o}$ dominates over that stored in internal stresses, $U_{m}$.  Because the optical potential is nearly lossless, the effects of material dissipation are diluted by a factor $U_{m}/(U_{o}+U_{m})$. Therefore, we would expect the mechanical $Q$ of the system to increase as the inverse of the dilution factor $\sim U_{o}/U_{m}$ for a frequency-independent damping mechanism.
For an ideal system, one would find an indefinite increase in the ratio $U_{o}/U_{m}=(f/f_0)^2$ with increasing trap power. In practice, this ratio saturates due to factors such as an inhomogeneous trapping beam profile and mode mixing with the modes of the support structure (tether and SiO$_2$ annulus in our case).  Accounting for these effects, our finite-element simulation predicts a maximum of $U_{o}/U_{m} \sim 100$ (see Fig. \ref{Qdata}(A)) in our experimentally achievable frequency range.

To infer the $Q$ of the trapped pendulum, we record the thermal motion $X(t)\propto V_{TB}(t)$ at each laser power for a few minutes and numerically compute the energy auto-correlation, $R_{E}(\tau) \equiv\langle  X(t)^2  X(t+\tau)^2\rangle$, over a Fourier frequency range encompassing the mechanical frequency~\cite{Stipe2001}. 
For a high $Q$ oscillator driven by Gaussian thermal noise, $R_{E}(\tau)$  is characterized by an exponential decay with time constant $\tau_{E}$, in correspondence to our measurements. The time constant $\tau_{E}$ relates to the mechanical quality factor by $Q=2\pi f \tau_E$. From shot to shot, we observe a 20\% standard deviation in the inferred value of $Q$. The pendulum in the absence of trapping forces has an initial value $Q_i=1.1(2)\times 10^4$. 

A summary of our $Q$-measurement results is presented in Fig. \ref{Qdata}(B).  Two datasets corresponding to slightly different optical alignments are compared.  In addition, we compare the result of monitoring vertical angular displacement of the pendulum on PD$_1$ (data in circles) with the result of monitoring axial displacement of the pendulum via the trap beam transmission through M$_2$ on PD$_2$ (triangles). The results are consistent in both cases. We observe that the initial $Q$ increase to a maximum value $Q_a=6.9(1.4)\times 10^5$ is consistent with the $(f/f_0)^2$ scaling and which is in contrast with the stress induced $Q$-increase studied in SiN nano-strings~\cite{Verbridge07}.
After the initial increase, $Q$  then drops as the CM mode crosses the violin mode, which is a consequence of an increase of the strain energy in the bending of the tether \cite{chang11}. Beyond the avoided crossing with the violin mode and a subsequent annulus mode (arrow $\times$), the $Q$ of the CM mode increases again by a factor of $>$ 50 relative to $Q_i$ to a final value $Q_f=5.8(1.1)\times 10^5$.
The dependence of $Q$ on frequency is in qualitative agreement with the calculated $U_{o}/U_{m}$ in Fig. \ref{Qdata}(A). Overall, we demonstrate that by adding optical energy but not dissipation into the system, we can increase the mechanical $Q$ by 
more than an order of magnitude.




We expect that further significant advances can be made with
refined fabrication techniques and a shift to materials
with better mechanical characteristics. For instance, the mechanical frequency 
and the corresponding $Q$-factor of our trapped pendulum is limited in part by 
the large suspended annulus to which the tether is attached (Fig. \ref{device}).  
Using wet chemical anisotropic etching of Si to release the pendulum, it should 
be possible to fabricate a device with an annulus less than 10 $\mu$m wide. 
Furthermore, while SiO$_2$ proved to be convenient to work with initially, it
suffers a relatively low intrinsic quality factor of $Q_i\sim
10^4$ that is likely to be limited by surface-related damping 
mechanisms~\cite{Yasumura2000, Penn2006}. 
Although the nature of surface damping is still an open question and 
not necessarily a fundamental limitation, we can still compare our observed $Q_i$
to other SiO$_2$ devices.
From the extensive phenomenological study
of SiO$_2$ loss angle~\cite{Penn2006} and the surface-to-volume ratio
of our pendulum, we would expect $Q_i\sim9200$, which is consistent with
our observation of $Q_i\approx1.1\times 10^4$. 
Switching platforms to stressed
silicon nitride or crystalline silicon should enable material quality
factors of $Q_i\sim 10^{5}$-$10^7$~\cite{Yang2000, verbridge06,wilson09}. In
initial experiments with Si$_3$N$_4$, for example, we have fabricated
stressed, tethered structures (similar to \cite{Kleckner11}) with bare frequencies of ${\sim}\,200$
kHz and $Q_i\sim 8\cdot10^6$. We expect that by applying optical
trapping to such structures, final quality factors of $Q_f\sim
10^8$ might be possible for oscillator frequency $\sim$ 1 MHz. 
Such values would be unprecedented for
any fabricated nano- or micro-mechanical system, and remarkably,
would be competitive with the prediction for untethered levitated
nano-particles~\cite{ashkin76,libbrecht04,chang10a,romero-isart10,li10}.

Our technique holds promise as a tool to reduce the role of mechanical dissipation in a wide variety of
sensing applications as well as in the emerging field of quantum
opto-mechanics~\cite{aspelmeyer10}. Our device can be integrated
into a high-finesse cavity employing the ``membrane-in-the-middle''
geometry~\cite{Thompson08,wilson09}, for example, and could provide the
long coherence times necessary to observe quantum behaviors
(i.e., macroscopic entanglement) in a room temperature environment~\cite{genes08b,muller-eberhardt08}. 
This work reveals  a fascinating new aspect of the interplay between motion 
and light~\cite{Braginsky} and should stimulate further exploration. 

We thank T. Rosenband for insightful discussions. 
KN acknowledges support from Caltech's Center for the Physics of Information (CPI). 
DEC acknowledges support from CPI and the Fundacio Cellex Barcelona.
This work is supported by 
DARPA ORCHID program, the NSF, DoD NSSEFF, and the 
Institute of Quantum Information and Matter, an NSF Physics Frontier Center 
with support of the Gordon and Betty Moore Foundation.

\bibliographystyle{apsrev}
\bibliography{bib_trapping}

\newpage

\newpage
\newpage

\end{document}